\documentclass[5p, times, twocolumn, preprint, authoryear]{elsarticle}
\usepackage[utf8]{inputenc}
\usepackage{graphicx}
\usepackage{booktabs}
\usepackage{url}
\usepackage{threeparttable}
\usepackage{tikz}

\def\mycolor{black}
\def\tablepicture#1{
  \begin{tikzpicture}[xscale=0.04, yscale=0.06]
    \useasboundingbox (-0.2,8) rectangle (16.2,5); 
    \begin{scope}[ycomb, yscale=0.95, xscale=5.5]
      \draw[fill=\mycolor!100!black!10, \mycolor!100!black!10] (0,0) 
rectangle (11,10); 
      \draw[yshift=100pt, \mycolor!70!black!80, line width=5] 
plot[] file {#1}; 
      \draw[\mycolor!100!black!25, line width=0.2] (-0.1,-0.4) rectangle 
(11.1,10.4);  
   \end{scope}
\end{tikzpicture}
}

\def\picture#1{
  \begin{tikzpicture}[xscale=0.04, yscale=0.04]
    \useasboundingbox (-2,2) rectangle (26,3.2); 
given just for demonstration.
    \begin{scope}[ycomb, xshift=-75pt, yscale=0.9, xscale=5]
      \draw[fill=\mycolor!100!black!10, \mycolor!100!black!10] (0,0) rectangle 
(6,10); 
      \draw[yshift=100pt, \mycolor!70!black!80, line width=5] plot[] file {#1}; 
      \draw[\mycolor!100!black!25, line width=0.2] (-0.1,-0.4) rectangle 
(6.1,10.4);  
   \end{scope}
\end{tikzpicture}
}

\begin{document}

\begin{frontmatter}

\title{Assessing Requirements Engineering and Software Test Alignment - Five 
Case Studies}
\author[bth]{Michael Unterkalmsteiner\corref{cor1}}
\ead{mun@bth.se}

\author[bth]{Tony Gorschek}
\ead{tgo@bth.se}

\author[bth]{Robert Feldt}
\ead{rfd@bth.se}

\author[bth]{Eriks Klotins}
\ead{ekx@bth.se}

\cortext[cor1]{Corresponding author. Tel.: +46 455 385815}
\address[bth]{Department of Software Engineering, Blekinge Institute of 
Technology, 371 79 Karlskrona, Sweden}

\begin{abstract}
The development of large, software-intensive systems is a complex undertaking 
that we generally tackle by a divide and conquer strategy. Companies thereby 
face the challenge of coordinating individual aspects of software 
development, in particular between requirements engineering (RE) and software 
testing (ST). A lack of REST alignment can not only lead to wasted effort but 
also to defective software.
However, before a company can improve the mechanisms of coordination they need 
to be understood first. With REST-bench we aim at providing an assessment tool 
that illustrates the coordination in software development projects and  
identify concrete improvement opportunities.
We have developed REST-bench on the sound fundamentals of a taxonomy on REST 
alignment methods and validated the method in five case studies. 
Following the principles of technical action research, we collaborated with 
five companies, applying REST-bench and iteratively improving the method based 
on the lessons we learned.
We applied REST-bench both in Agile and plan-driven environments, in projects 
lasting from weeks to years, and staffed as large as 1000 employees. The 
improvement opportunities we identified and the feedback we received indicate 
that the assessment was effective and efficient. Furthermore, participants 
confirmed that their understanding on the coordination between RE and ST 
improved.
\end{abstract}


\end{frontmatter}

\section{Introduction}
Requirements Engineering (RE) is the discipline of eliciting, analyzing, 
specifying, validating and managing needs and constraints on a software 
product~\citep{nuseibeh_requirements_2000, bourque_guide_2014}. 
Software Testing (ST) is the verification that a software product provides 
expected behaviors, as expressed in 
requirements~\citep{bertolino_software_2007, bourque_guide_2014}. As such, RE 
and ST\footnote{We abbreviate requirements engineering and software testing as 
``RE and ST'' or ``REST'' in the remainder of this paper.} are intrinsically 
related and leveraging on this relationship would be beneficial for both 
disciplines~\citep{graham_requirements_2002}. \cite{tassey_economic_2002} 
projected the cost of inadequate testing in the US to 60 billion dollars per 
year. Furthermore, a survey by \cite{garousi_replicated_2010} found that 
defects introduced in the requirements were among the most expensive to 
repair (besides stress\slash performance problems). 
\cite{bjarnason_challenges_2014} identified sixteen REST alignment challenges, 
spanning from requirements and test quality to requirements abstraction 
levels~\citep{gorschek_model_2006} and traceability.

First steps to a better understanding of the REST alignment phenomenon were 
undertaken by studying and classifying alignment 
practices~\citep{unterkalmsteiner_taxonomy_2014}. The main contribution of 
this classification is the definition of an epistemic 
base~\citep{mokyr_chapter_2005} that can be used to explain how and why REST 
alignment practices work. In this paper we use this base in order to provide a 
practical method, REST-bench, to assess REST alignment. A prerequisite for any 
improvement is the characterization of the current condition of the phenomenon 
under study. Based on this agreed state and the definition of goals, changes 
can be designed and implemented. Postmortems~\citep{birk_postmortem:_2002} are 
one possibility to elicit best practices but also issues in the execution of 
projects, feeding the results into an organizational knowledge 
repository~\citep{ivarsson_tool_2012}. Even though guidelines for executing 
postmortems exist~\citep{collier_defined_1996,dingsoyr_postmortem_2005}, 
postmortem reviews are seldom held, some suggest for lack of 
time~\citep{keegan_quantity_2001,glass_project_2002}, even though their 
benefits are well reported~\citep{verner_-house_2005}. REST-bench follows, like 
postmortems, the principle of the Experience 
Factory~\citep{basili_improve_1995}, where improvements are based on 
data collection and analysis of experience from past projects.

We designed REST-bench to be lightweight, in terms of resource use (30-50 
person-hours per application), by focusing the assessment effort to the 
specific issue of requirements engineering and test coordination which is of 
major interest to organizations developing software intensive 
systems~\citep{bjarnason_challenges_2014}. REST-bench is 
interview- and workshop-based, providing structured guidelines to collect and 
analyze data originating from project participants. In essence, REST-bench 
illustrates the impact of project artifacts on the coordination between RE and 
ST, and provides a set of analytical tools (the artifact map, seeding 
questions) that drive the analysis and elaboration of improvement suggestions.

In this paper, we present REST-bench in an example-driven manner, describing 
data elicitation, data preparation and the collaborative analysis. We 
illustrate the application of REST-bench in five companies that exhibit diverse 
characteristics. In all five cases we could identify relevant improvement 
opportunities. The participants of the assessment judged REST-bench as an 
efficient and effective mean to assess the coordination between RE and ST. 

The remainder of this paper is structured as follows. We discuss background and 
related work in Section~\ref{sec:background}. Section~\ref{sec:rm} illustrates 
the research method we followed to validate and improve REST-bench. We 
introduce the assessment method in Section~\ref{sec:rb}, together with a 
running example that shows the application of REST-bench and with the 
improvements we implemented, the collected data, the analysis of the results 
and the identified improvement potential. In Section~\ref{sec:casestudies} we 
show the results of the remaining four case studies. We answer our initially 
stated research questions in Section~\ref{sec:discussion} and conclude the 
paper in Section~\ref{sec:conclusions}.

\section{Background and Related Work}\label{sec:background}
\begin{figure}
	\centering
	\includegraphics[width=0.98\linewidth]{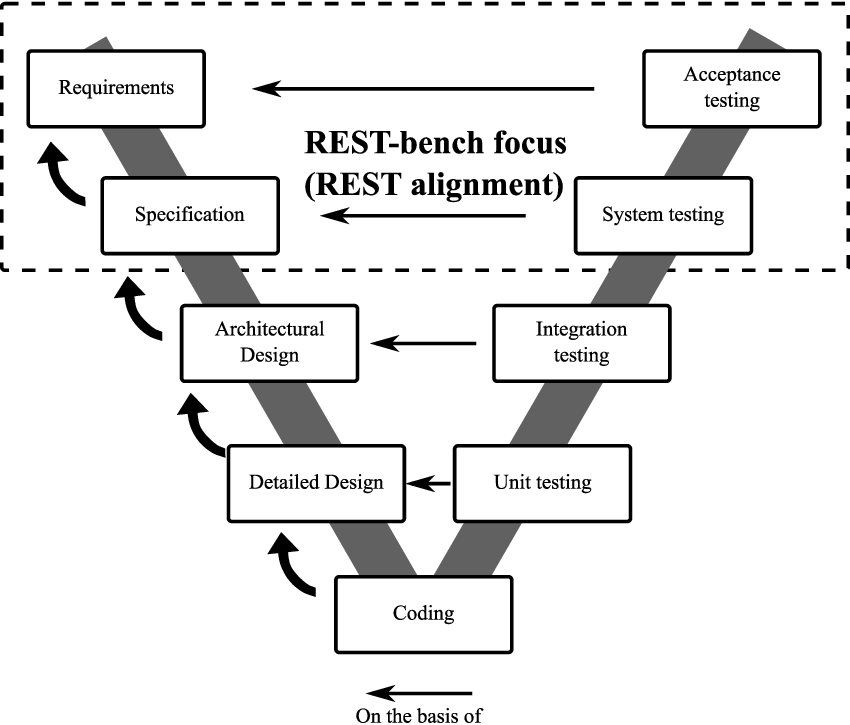}
	\caption{V-Model of software development}
	\label{fig:vmodel}
\end{figure}
\subsection{REST alignment}
The development of software-intensive systems\footnote{A software-intensive 
system is ``any system where software contributes essential influences to the 
design, construction, deployment, and evolution of the system as a 
whole''~\citep{iso/iec_iso/iec_2007}.} is a collaborative effort of 
experts, each contributing a part to the solution~\citep{rus_knowledge_2002}. 
Expertise and capabilities are distributed on different roles, rendering the 
coordination between people in software projects and development phases 
essential for success~\citep{kraut_coordination_1995}. From a process 
perspective, software development consists of transitions
from system concept, requirements specification, analysis and design, 
implementation, and test and maintenance~\citep{laplante_what_2007}. This 
abstraction holds for both plan-driven process models (e.g. 
spiral~\citep{boehm_spiral_1988} and 
evolutionary~\citep{naumann_prototyping:_1982}, and the unified process 
model~\citep{kruchten_rational_2000}, as well as Agile models, although to a 
lesser extent as activities may be blended, eliminating transitions altogether 
(e.g. in eXtreme Programming~\citep{beck_embracing_1999}).

Figure~\ref{fig:vmodel} illustrates the V-Model of software development, which 
originates from system engineering~\citep{forsberg_relationship_1991, 
brohl_v-_1995} and was adopted in software 
engineering~\citep{pfleeger_software_2009}. This model illustrates 
vertical transitions between software development activities (left hand side) 
and the corresponding testing activities (right hand side) that ought to ensure 
the quality of the resulting work products. The importance of enabling these 
vertical transitions and align the intentions and activities across is 
demonstrated by an abundance of research (e.g. between requirements abstraction 
levels~\citep{gorschek_requirements_2006}, requirements and software 
architecture\slash design~\citep{kop_conceptual_1998, amyot_bridging_2001, 
hall_relating_2002}, software architecture\slash design and 
implementation~\citep{murphy_software_2001, elrad_aspect-oriented_2002, 
aldrich_archjava:_2002}, and software architecture\slash design and 
testing~\citep{muccini_using_2004, samuel_automatic_2007}).

While research has produced an ample amount of software technologies, models 
and frameworks to coordinate people, to ease the transition between software 
development phases and to align the intentions and activities therein, 
literature is sparse on methods that focus on improving the coordination 
between requirements engineering and software 
testing~\citep{bjarnason_challenges_2014}. In earlier 
work~\citep{unterkalmsteiner_taxonomy_2014} we defined REST alignment 
as the \emph{adjustment of RE and ST efforts for coordinated functioning and 
optimized product development}. The key in this definition is the intuition 
that RE and ST efforts need to be adjusted together in order to avoid 
sub-optimization of either one of the two aspects. This adjustment can be 
achieved by various means, spanning from process-centered activities that 
foster the collaboration between RE and ST roles (e.g. by forming 
cross-functional teams~\citep{marczak_how_2011}), over techniques that use 
requirements as a driver for testing activities (e.g. by formulating testable 
contracts~\citep{melnik_executable_2006} or model-based 
testing~\citep{utting_taxonomy_2012}), to methods or processes that 
establish and\slash or maintain requirements to test traceability 
links~\citep{ramesh_toward_2001}. Fundamental for any form of coordination is 
the exchange of information~\citep{van_de_ven_determinants_1976}. Therefore, in 
order to characterize the means of REST alignment, we use information as a 
central entity, as described next.

\subsection{The REST taxonomy}
\begin{figure}
\centering
\includegraphics[width=0.98\linewidth]{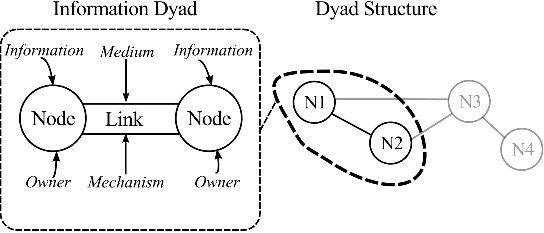}
\caption{Information dyad and dyad structure}
\label{fig:dyad_structure}
\end{figure}

\begin{table}[t]
\footnotesize
\caption{Dyad structure properties}\label{tab:dsp}
\begin{tabular}{lp{0.86\columnwidth}}\toprule
\emph{P1} & Number of nodes -- links between nodes need to be maintained over 
	time. Hence, the total number of nodes allows one to reason on complexity 
	and effort to maintain REST alignment.\\
\emph{P2} & Branches -- a branch exists, if a node acts as a source or sink for 
	more than one node. Branches may reduce local complexity, require however 
	also means to synchronize and merge information. \\
\emph{P3} & Intermediate nodes -- characterized by information that belongs to 
	the design\slash analysis or implementation software development 
	phases\slash activities.\\
\emph{P4} & RE and ST node proportion -- assuming that a node is associated 
	with a certain cost (e.g. establishing\slash maintaining the information 
	therein and links in between), it is of interest to know the node 
	distribution among the RE and ST software development phases\slash 
	activities. \\
\emph{P5} & Links -- the linking mechanism between software development 
phases\slash activities determines how changes of information are propagated. \\
\emph{P6} & Scope -- allows one to reason upon the interface between RE and ST 
	and other software development phases\slash activities.\\
\bottomrule
\end{tabular}
\end{table}

To characterize means aimed at achieving REST alignment, we developed a 
taxonomy that uses the information dyad as building block to describe alignment 
methods~\citep{unterkalmsteiner_taxonomy_2014}. Figure~\ref{fig:dyad_structure} 
(left) illustrates the components of an information dyad: two nodes, 
characterized by the information they contain and their owner, are connected 
trough a link. Creating the REST taxonomy, we identified different linking 
media and mechanisms. A medium can be of several different types: structured 
artifacts (e.g. documents, email, diagrams, database records); unstructured 
artifacts (e.g. audio and video); tools that act as means to share, transfer or 
transform information (e.g. modeling tools, language analysis tools); processes 
(one or more activities, performed repeatedly); or the organization of work 
environment (co-location, role\slash responsibility rotation). Furthermore, we
identified four types of linking mechanisms: implicit connection (information 
is connected by volatile and implicit links that are not formalized); 
connection (information pertaining in each node is connected, establishing a 
logical link between the two nodes); bridge (information pertaining to each 
node is connected and augmented in order to achieve fitness of purpose in both 
nodes); and transformation (information, packaged for one node in the alignment 
dyad, is re-packaged in order to satisfy the needs of the other node). 

Information dyads form a structure, Figure~\ref{fig:dyad_structure} (right), 
which in turn may exhibit certain properties, listed in Table~\ref{tab:dsp}. 
We have used these properties to compare REST alignment methods reported in 
literature~\citep{unterkalmsteiner_taxonomy_2014}. In this paper, one objective 
is to study to what extent these properties support the identification of 
improvement opportunities for REST alignment. Concretely, we used the dyad 
structure properties to generate seedings questions that support the analysis 
of the collected data in the REST-bench method. The steps in this process and 
the used seeding questions are illustrated in Section~\ref{sec:rb}, while the 
question whether all dyad structure properties were useful is answered in 
Section~\ref{sec:discussion}. 

\subsection{Related work}
\begin{table*}
	\footnotesize
	\caption{Project characteristics of the participating 
	companies}\label{tab:companies}
	\center
	\begin{tabular}{p{2.5cm}p{3cm}lp{3cm}p{3cm}p{2cm}} \toprule
		Company & A & B & C & D & E \\ \midrule
		Project duration & 12 months & 2 months/release & 36 months & 48 months 
		& 7 
		months \\
		Staff & 150 & 20 & 9 & 1000 & 20 \\
		Software development approach & Agile (in transition from plan-driven) 
		& Agile & Agile (embedded in plan-driven) & plan-driven & plan-driven \\
		Requirements \# & 350 & 5-20 & 300 & 2000 & 50\\
		Test-case \# & 700 & Not stated & 300 & 500 & 24\\ 
		Assessment date & 2012/06 & 2013/03 & 2013/03 & 2013/04 & 2013/12 \\ 
		\bottomrule
	\end{tabular}
\end{table*}
The challenges of coordinating 
development teams operating at different sites have been described 
by \cite{herbsleb_splitting_1999}. They point out the boundaries of explicit 
coordination mechanisms such as plans, interface specifications and process 
descriptions, but also the lack of informal coordination opportunities in 
geographically distributed sites, leading to misunderstandings and increase in 
cycle time for fixing issues. Following this line of thought, 
\cite{herbsleb_formulation_2003} developed an empirical theory of coordination 
(ETC) for software engineering, based on decision and constraint networks, 
and later applied to identify coordination requirements among software  
developers that can be used to improve the design of collaboration 
tools~\citep{cataldo_identification_2006}. Along a similar line, 
\cite{ko_information_2007} observed in a field study what information 
developers seek in their day-to-day work, which sources they use and what the 
reasons are for not being able to gather the needed information, calling for 
innovation in tools, processes and notations. REST-bench shares with these 
studies the insight that satisfying information needs, timely and as exhaustive 
as possible, is key to successful software development in teams. 

In order to represent information flow in requirements engineering activities, 
\cite{schneider_beyond_2008} developed a notation that can be used to model 
both formal and informal communication. A benefit of the flow notation is that 
it can be used to describe the officially required and the actually executed 
process, showing differences and instances where improvements can be 
implemented~\citep{stapel_improving_2007}. Furthermore, FLOW mapping has been 
used to plan and manage communication in distributed 
teams, requires however a considerable amount of manual work to keep the maps 
up-to-date with continuously changing communication 
patterns~\citep{stapel_flow_2011}. REST-bench shares with FLOW the idea to 
represent information and connections in a diagram that can be discussed in 
collaboration with practitioners to identify improvement opportunities. On the 
other hand, REST-bench provides a concrete assessment process, differentiates 
between data collected from the different roles, and provides heuristics that 
can be used to analyze the collected data and to generate analysis points that 
may lead to improvement suggestions.

Project postmortems have been successfully applied in the 
past (\cite{dingsoyr_postmortem_2005} provides an overview and examples) to 
identify improvements, following the Experience Factory 
principle~\citep{basili_improve_1995}. However, \cite{glass_project_2002} has 
observed that they are seldom conducted due to the fast paced nature of 
software projects where teams are split up and reassigned to new tasks. 
Therefore, collecting information timely after the conclusion of a project 
seems more likely to identify improvement potential. In order to increase the 
data accuracy, \cite{bjarnason_evidence-based_2012} propose project timelines 
that are prepared in advance. Depending on the particular analysis goals, 
certain aspects of the collected data are visualized in a timeline, allowing 
the postmortem participants to recall the illustrated events. However, since 
the method can be used for any generic improvement goal, it does not provide 
prompting questions or aids that could facilitate the analysis of the collected 
data. While REST-bench relies also on the collection of data from a specific 
past project, eliciting practitioners' experience on how information is used 
and created, it provides also seeding questions (Table~\ref{tab:questions}) 
that support the analysis or the collected data. The focus on a particular goal 
(coordination of RE and ST), the structured data collection, and the analysis 
guided by a set of predefined questions and an artifact map, sets REST-bench 
apart from traditional project postmortems, as for 
example described by \cite{dingsoyr_postmortem_2005}.

\section{Research Method}\label{sec:rm}
Our overall research approach is oriented towards design 
research~\citep{hevner_design_2004} which provides a concrete framework for 
implementing the dynamic validation phase in the technology transfer model 
proposed by~\cite{gorschek_model_2006}. Our research method is best described 
as technical action research~\citep[Ch. 19]{wieringa_design_2014}, i.e. we aim 
at improving and validating the fitness of purpose of an artifact by applying 
it in a real-world environment~\citep{wieringa_empirical_2014}. In particular, 
we want to answer the following research questions:
\begin{description}
 \item RQ1: To what extent are the dyad structures from the 
REST taxonomy useful to elicit improvement opportunities?
 \item RQ2: To what extent is REST-bench useful in Agile and plan-driven
environments?
  \item RQ3: To what extent is REST-bench usable?
\end{description}

In our previous work, we characterized REST alignment methods by means of 
information dyads~\citep{unterkalmsteiner_taxonomy_2014}. Furthermore, we 
piloted the idea of using dyad structures as a mean to drive REST alignment 
assessment. With RQ1 we aim to validate this idea by applying the REST-bench 
method in a series of case studies. When we planned the validation, one of our 
concerns was the methods' reliance on documentation as a proxy to determine 
REST alignment. Therefore, we questioned whether we can apply REST-bench at all 
in an Agile environment. We address this concern in RQ2, where a 
plan-driven environment means that the development teams work in a traditional, 
document driven manner~\citep{petersen_effect_2010}. 
In order to be adopted by practitioners, a method should also be usable. We 
analyze the usability of REST-bench from the perspective of the analyst, i.e. 
the researcher who applied the method, as well as through the practitioners' 
feedback who participated in the study.

\subsection{Project characteristics}
We applied REST-bench in five companies located in Sweden: Ericsson and Telenor 
in Karlskrona, ST Ericsson in Lund, CompuGroup Medical and Volvo Cars in 
Gothenburg.
All companies were approached based on personal contacts, providing them an 
executive summary of the goals, and expected cost and benefits of REST-bench. 
We did not preclude any particular company or project type in the selection. In 
the remainder of this paper we anonymized project characteristics, collected 
data and results by referring to Company A, B, C, D, and E.
Table~\ref{tab:companies} illustrates the characteristics of the projects where 
we applied REST-bench. Note that Company A was, at the time of 
the assessment, in a transition from a plan-driven to an Agile software 
development approach. The team in Company C was working in an Agile manner 
while still being embedded in a plan-driven process. We selected the particular 
projects based on the criteria defined by REST-bench, described in 
Section~\ref{sec:rb-selection}.

\begin{figure*}
	\centering
	\includegraphics[width=\textwidth]{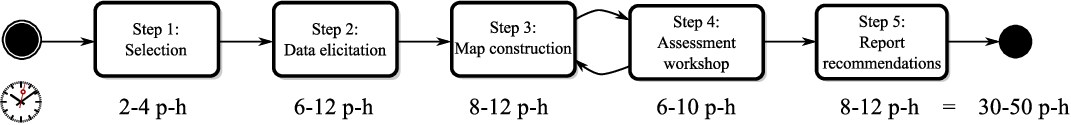}
	\caption{The steps in the REST-bench method and budgeted 
		effort in person-hours (p-h)}\label{fig:restbench}
\end{figure*}  

\subsection{Data collection and analysis}
We collected data from two sources: (1) the process of applying REST-bench, 
e.g. effort spent and identified useful seeding questions; (2) a dedicated 
questionnaire, distributed to the participants after the assessment.
Table~\ref{tab:questionaire} shows the 10 questions. The first four questions 
are open ended, addressing the overall experience of using REST-bench. The 
remaining questions required an answer on a 5 point Likert scale (strongly 
agree, agree, neutral, disagree, strongly disagree), while explanatory feedback 
was still possible. Out of the 13 participants in total, 10 returned the 
questionnaire.

In order to answer RQ1, we collected the seeding questions (see 
Table~\ref{tab:questions}) used in the five collaborative workshops and 
analyzed their association with the dyad structure properties shown in 
Table~\ref{tab:dsp}. Furthermore, we use answers from the questionnaire to 
provide evidence for or against the relevance and usefulness of REST-bench.

Data to answer RQ2 stems from the project characteristics of the participating 
companies (Table~\ref{tab:companies}), allowing us to stratify the results from 
the questionnaire into plan-driven/Agile sets. Furthermore, the assessment 
results from plan-driven and Agile environments provide indications on the 
usefulness of the approach.

In order to answer RQ3, we use data from the questionnaire, in particular 
questions regarding efficiency and effectiveness of the approach, and collect 
data on effort spent for the different steps in REST-bench.

\subsection{Limitations}
A major threat to validity in any technical action research is the involvement 
of the researcher in the application of the to be validated 
artifact~\citep{wieringa_technical_2012}. The question whether the validation 
results depend solely on the ability of the researcher to use the artifact 
cannot be answered conclusively. In this paper, the first author performed all 
steps in REST-bench. An alternative would have been to train practitioners and 
to let them apply REST-bench on their own, observing the application and 
collecting data of spent effort, usefulness and usability. However, this would 
have led to a considerably higher assessment cost per company. Nevertheless, 
the validation did not solely depend on the researcher since practitioners 
where involved in the REST-bench application process, influencing thereby 
considerably the results. 

We applied convenience sampling~\citep{robson_real_2002} to identify the case 
companies, i.e. we approached engineers and managers we had collaborated in the 
past. We prepared information material on REST-bench (purpose, expected 
effort, outcome and deliverables), providing the contact persons in the 
companies a basis on which they can decide whether or not to engage (all 
approached companies did). Note however that the selection of projects and 
interviewees followed the requirements of Step 1 of the REST-bench method (see 
Section~\ref{sec:rb-selection}). We have no indications that the contact 
persons at the companies influenced, other than by supporting the 
identification of interviewees, the outcome of the assessment with REST-bench 
or the questionnaire results. 

We applied REST-bench in projects with a diverse set of characteristics (see 
Table~\ref{tab:companies}). There is no feature or constraint in REST-bench 
that could conceivably exclude certain classes of projects or companies, even 
though very small teams that work co-located and do not require coordination 
with other teams might not benefit from an assessment with REST-bench. Note 
that REST-bench does \emph{not} assess coordination through oral and informal 
communication which are still central for the successful execution of a 
software project (see answers to Q10 shown in Table~\ref{tab:questionaire}).
The post-assessment questionnaire was returned by 10 (out of 13) participants. 
While this is a high relative response rate (77\%), the evidence provided by 
the questionnaire, in absolute terms, is still weak.

\section{REST-bench}\label{sec:rb}
REST-bench follows the macro procedure of lightweight software 
process assessment~\citep{pettersson_practitioners_2008}, however it has a 
focused improvement goal (coordination between RE and ST) and utilizes 
elicitation and analysis practices tailored to this goal. This focus leads to a 
low investment cost, typically between 30 and 50 person-hours (p-h), depending 
on how many people are involved in the assessment (typically 3 to 5). We have a 
project view to make the assessment concrete, and avoid people give us the best 
or worst cases picked together from any part at any time in the company. This 
project focused assessment has been used before with 
success~\citep{svahnberg_uni-repm:_2015}.

The aim of REST-bench is to assess the state of REST 
alignment~\citep{unterkalmsteiner_taxonomy_2014}, eventually initiating changes 
to improve that alignment. A critical enabler for any kind of change in an 
organization is support by senior 
management~\citep{dyba_empirical_2005,niazi_critical_2006,wohlin_success_2012}. 
Therefore, we regard a champion in the organization that can provide resources 
and support the dissemination of the assessment results an essential 
pre-requisite before embarking in a REST-bench assessment. 

We performed a REST-bench assessment at five case companies, and used the 
experiences to refine REST-bench. The first four cases (A-D) are 
illustrated in chronological order in Section~\ref{sec:casestudies}, where we 
show the results of the assessment and the adaptions we made to the method. In 
this section we use the fifth case (E) as a running example to illustrate the 
individual steps of REST-bench. In all five case illustrations, the first 
author of this paper acted as moderator and analyst during the assessment.

REST-bench is interview-driven and begins therefore with the 
selection of interviewees (Step 1 in Figure~\ref{fig:restbench}). After data 
elicitation (Step 2), the analyst creates an artifact map (Step 3). In Step 4, 
the analyst meets with the interviewees to collaboratively identify improvement 
opportunities, using the artifact map as input. The results of this assessment 
workshop are collected in a report (Step 5). In 
subsections~\ref{sec:rb-selection}~-~\ref{sec:rb-recommendations} 
we exemplify the steps shown in Figure~\ref{fig:restbench}, providing effort 
estimation and a compilation of best practices for the application of 
REST-bench.

\subsection{Step 1 -- Selection}\label{sec:rb-selection}
REST-bench has an interview-driven data elicitation process. Therefore, the 
selection of the particular interviewees is pivotal for the assessment, not 
unlike most process assessment methods such as e.g. CMMI 
SCAMPI~\citep{sei_appraisal_2006} or SPICE~\citep{rout_spice_2007}. A 
local champion in the organization, supporting the initiative, can be a great 
accelerator in identifying interview candidates. In order to maintain 
REST-bench's lightweight and focused nature, the roles of the 2-4 interviewees 
should pertain to the requirements domain, e.g. requirements engineers, 
business analysts, product managers, and the testing domain, e.g. test and 
quality assurance engineers. We recommend that the selection process considers 
the following characteristics:
\begin{enumerate}
 \item Work experience in the company and in the particular role. A candidate 
that knows ''how things work`` and has progressed in the same company to a 
senior position is preferable over a qualified, but new hire.  
 \item The candidate from the RE and ST domain must have collaborated on the 
same project, preferably in more than one instance.
 \item The project has recently been closed or is in a late stage.
\end{enumerate}

These characteristics allow us to elicit data that reflect how engineers 
actually work as opposed to how they ought to work according to a process 
prescribed by a company policy. Choosing a project on which all interviewees 
collaborated allows us to conduct episodic 
interviews which facilitates the collection of ''everyday knowledge about 
objects or processes``~\citep[p. 85]{bauer_qualitative_2000}. In episodic 
interviews, we try use actual events that are connected to actions, experiences 
and consequences, to elicit relevant information from project 
participants. A project that is in its early stage can not set the context as 
this everyday knowledge (episodes) has still to be acquired. We detail data 
elicitation in Section~\ref{sec:elicitation}.

In some organizations, the notion of a ''project`` is not clear cut. For 
example, one case company applies SCRUM~\citep{schwaber_scrum_1997} with three 
week sprints to implement customer requirements, releasing a new version of the 
product every two sprints. In this situation we chose the last sprint as 
''project`` scope for the interviews. The relevant decision here is to agree on 
a specific context such that all interviewees can relate to it.

When selecting the interviewees it is important to keep the main goal of the 
assessment in mind: identify improvement opportunities for the coordination 
between requirements engineering and testing aspects of software development. 
Optimal candidates are employees in a senior position, being however 
involved in the day-to-day practice in their respective area. This leads 
usually to very fruitful analysis during the collaborative issue identification 
(Section~\ref{sec:collab}).

\subsubsection*{Effort and Best Practices}
The budgeted effort for this step is 2-4 person-hours, where the majority of 
the expense is on the organization, identifying the interview candidates. The 
analyst can support this by providing detailed selection criteria for projects 
and interviewees, described in this section. We recommend to schedule all 
meetings in advance (interviews, workshop) such that the overall assessment 
procedure is not extended to a long time period. Interviews should be held on 
one day, increasing the efficiency of data collection. The assessment workshop 
should be scheduled between 4 and 10 business days after the interviews, 
allowing time to analyze the collected data and preventing a loss of context by 
waiting too long after data collection.

\subsubsection*{Example - Step 1}
Company E outsources the implementation and verification of parts of their 
product to an external supplier. In this scenario, the coordination between RE 
(Company E) and ST (supplier) is particularly challenging since company, 
country and time-zone borders are crossed. The selection of interviewee 
candidates was supported by the company's responsible for process improvement. 
We organized a seminar for employees of Company E, presenting the goals and 
example applications of REST-bench. This pro-activeness raised interest in the 
method and helped us to identify the project and interviewees for the 
assessment. We chose an Analyst Lead and a Business Process Expert/Acceptance 
Tester for the RE perspective, both working for Company E. For the ST 
perspective, we chose an Acceptance Test Manager working at Company E and the 
System Test Manager working for the supplier. The chosen project had a duration 
of 7 months involving a total staff of 20 employees (both Company E and 
supplier). 

\subsection{Step 2 -- Data elicitation}\label{sec:elicitation}
We argued in our earlier work on requirements engineering and software testing 
alignment methods~\citep{unterkalmsteiner_taxonomy_2014} that the 
means of connecting and using information is central to any software 
development effort. In order to operationalize this concept in a data collection 
procedure, we elicit which artifacts requirements and test engineers use and 
create in their daily work. We chose artifacts as a tangible proxy for 
information which can be retrieved during an interview.
Furthermore, every software organization produces some set of artifacts, 
independently how agile or lean the organization is. An artifact can be any 
kind of digital or analog document produced by an employee. In our interview 
guide we exemplify what we mean by artifacts (see  
Table~\ref{tab:exampleartifacts} for a list of examples).

\begin{table}
 \footnotesize
 \caption{Example artifacts}\label{tab:exampleartifacts}
 \begin{tabular}{p{0.95\columnwidth}}\toprule
 Powerpoint presentations, spreadsheets, text documents, specification / use 
case / user story stored in requirements management tool, acceptance / 
integration, system, unit test case stored in test management tool, UML / 
entity-relationship diagrams, emails, meeting notes, yellow sticky notes \\
 \bottomrule
 \end{tabular}
\end{table}

Agreeing on the concept of 
artifacts is key to the episodic interview technique as it sets the scope of 
what data is elicited. Therefore, Phase 1\footnote{In Phase 0, we ask a set of 
introductory questions to elicit context information, such as name of the 
project, project duration, staff size, applied software development method, 
number of system requirements and test cases.} requires the interviewee to list 
all 
artifacts he/she has used or created in his/her daily work. The interviewee 
should follow the timeline of the project under investigation, recalling 
his/her involvement during the different project phases that required use or 
creation of any artifact. In Phase 2 of the interview we elicit the following 
data on each of the artifacts:
\begin{itemize}
 \item Purpose: content of the artifact and reason for its creation
 \item Creator: role of the person who creates the artifact, in which 
phase of the project 
 \item User: role of the person using the artifact, in which phase of the 
project
 \item Modifier: role of the person changing the artifact, in which phase of 
the project
 \item Link or Mapping: a link is a uni-directional connection from one 
artifact to another; a mapping is a bi-directional connection between 
information contained in two artifacts.
 \item Use: a reference to any other artifact that is used as input to create 
this artifact
\end{itemize}

We collect this data by compiling a simple template (see 
Figure~\ref{fig:elicitationexample}). It is common that the list elicited in 
Phase 1 is incomplete and is extended while detailing the artifact 
information in Phase 2 of the interview. 

In order to reduce mutual influence during data collection, we interview 
requirements and test engineers separately. We also recommend to audio record 
the interviews (provided the interviewees give consent). With the episodic 
interview technique, we don't elicit data on artifacts in a vacuum, but enrich 
the information by relating it to the studied project and what has been 
experienced in the use and creation of artifacts. The analyst can then, while 
constructing the artifact map (see Section~\ref{sec:mapconstruction}), develop 
a better understanding of the coordination between RE and ST, and prepare 
targeted analysis points for the assessment workshop.

\subsubsection*{Effort and Best Practices}
The budgeted effort for this step is 6-12 person-hours. We allocate 1.5 hours 
per interview, which translates into 6 person-hours for 1 analyst and 2 
interviewees as we interview RE and ST separately. Adding two interviewees 
doubles the effort to 12 person-hours. 

Interviewees should have access to the project documentation during the data 
elicitation. This allows them not only to exemplify the artifacts they mention, 
but also to provide more accurate information on how artifacts are linked and 
related to each other.

\subsubsection*{Example - Step 2}
\begin{figure}[t]
 \centering
 \includegraphics[width=\columnwidth]{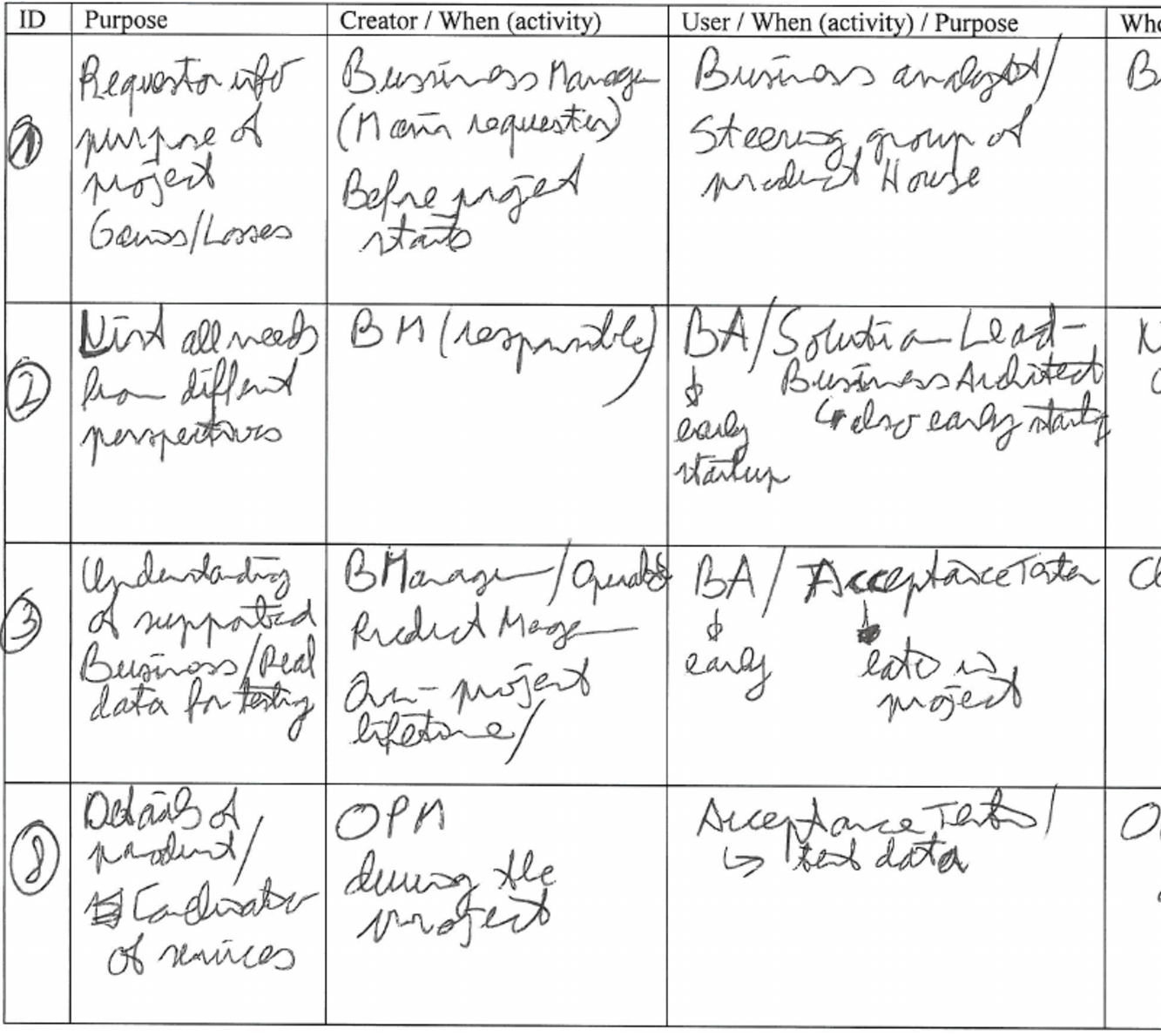}
 \caption{Artifact elicitation example}\label{fig:elicitationexample}
\end{figure}

The three interviews at Company E were performed on two days. The ST 
representative from the supplier company was interviewed by phone, however not 
audio recorded upon request by the interviewee. 
Figure~\ref{fig:elicitationexample} shows an excerpt of the simple elicitation 
form used to record data. We experienced that notes are sufficient to record 
facts about artifacts, eliciting them in a structured fashion. However, 
episodic information on their use or misuse is complex and requires audio 
recordings that can be analyzed offline. When interviewees refer to events in 
the project, the analyst should focus on understanding these occurrences, in 
order to pose follow-up questions, rather than spending his attention on 
recording them manually.

\subsection{Step 3 -- Map construction}\label{sec:mapconstruction}
\begin{figure}[t]
 \centering
 \includegraphics[width=0.8\columnwidth]{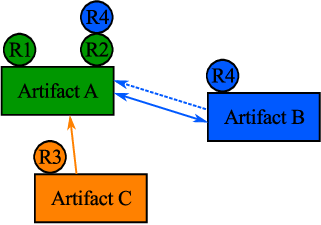}
 \caption{Artifact map - Illustration of components}\label{fig:mapexample}
\end{figure}
An artifact map is an easy to understand, graphical representation of the data 
collected during the interviews. The visualization serves mainly three purposes:
\begin{enumerate}
 \item Illustrating commonalities and differences of the RE and ST perspective 
on use and creation of artifacts.
 \item Generating input for the collaborative issue identification.
 \item Communicating and displaying the artifact topology to employees not 
involved in the assessment.
\end{enumerate}

Figure~\ref{fig:mapexample} illustrates the components of an artifact map. A 
rectangular box represents an artifact, a sphere with an identifier indicates a 
creator (on the top-left of the box) or an user (on the top-right of the box) 
of an artifact. ''Linked-to`` relationships are illustrated by solid lines with 
a single arrow, meaning that one artifact is linked to another. ''Mapped-to`` 
relationships are illustrated with double arrows, meaning that the information 
in the artifacts is mapped bi-directionally. ''Used-to-create`` relationships 
are illustrated by dashed lines with a single arrow, meaning that information 
in one artifact is used to create another artifact. The color coding indicates 
whether the data stems from RE (orange), ST (blue) or from both 
roles (green). The map in Figure~\ref{fig:mapexample} reads as follows: 
\emph{RE and ST agree that Artifact A is created by R1 and used by R2, even 
though ST adds R4 as an user. RE states that Artifact C is created by R3 and is 
linked to Artifact A. ST states that Artifact B is created by R4, using 
Artifact A is input, and information in Artifact B is mapped to information in 
Artifact A.}

This example illustrates a rather common situation where ST states that a role 
(R4) uses information from one artifact (A) to create another artifact (B), 
whereas RE is not aware of this information need. This leads to an analysis 
point that needs to be addressed in the collaborative assessment workshop. One 
outcome could be that missing to mention Artifact B and R4 was an oversight by 
RE. Then the map is simply updated. Another potential outcome is that RE is 
indeed unaware of the information need by R4. This misalignment could then be 
analyzed in more depth during the assessment workshop, identifying potential 
causes but also consequences. 
For example, if Artifact A is seldom updated, R4 might use outdated 
information to create Artifact B. The solution could be to maintain Artifact A 
during the project. Another likely solution: R4 should use Artifact C which 
would otherwise have no use (no user was defined for it during the 
elicitation). 

Table~\ref{tab:questions} lists questions that the analyst can use to identify 
analysis points for the assessment workshop. The questions in set \emph{P0} 
focus on 
finding causes of disagreement between RE and ST, which could be a 
misunderstanding in data elicitation, but also a genuine divergence that 
potentially causes misalignment. The questions in sets \emph{P1, P2, P5 and P6} 
are mapped to the information dyad structure properties identified in our 
previous work~\citep{unterkalmsteiner_taxonomy_2014}. Note that these questions 
are meant as an inspiration for the analyst and do not apply to every artifact 
map. While applying the REST-bench method, we improved the formulation and 
extended the set of questions. 

An artifact map encodes only a subset of the information elicited during the 
interviews. It is the task of the 
analyst to use the remaining data together with the collected episodic 
information to create analysis points for the collaborative assessment 
workshop. For example, the data elicitation may reveal that an artifact is 
regularly updated during a project. However, there is no mechanism to propagate 
changes to artifacts or roles that use this modified information, leading 
potentially to misalignment. Episodic information, such as an event where an ST 
had to redesign a test suite due to outdated requirements specifications can be 
supporting evidence for a pattern observed in the artifact map.

\begin{table}[t]
 \footnotesize
 \caption{Seeding questions to prepare the REST-bench assessment 
 workshop}\label{tab:questions}
 \begin{tabular}{p{0.95\columnwidth}}\toprule
  \multicolumn{1}{c}{\emph{P0 - What is the source of disagreement between RE 
and ST on...}} \\
  1. ... the existence of an artifact? \\
  2. ... the creator/user of an artifact? \\
  3. ... who changes when an artifact? \\
  4. ... ''linked-to``/''mapped-to`` relationships between artifacts? \\
  5. ... linking mechanism between artifacts? \\
  6. ... ''used-to-create`` relationships between artifacts? \\
  \midrule
  \multicolumn{1}{c}{\emph{P1 - Number of artifacts}} \\
  6. Is there an information need that was not fulfilled by the used artifacts? 
\\
  7. If a new artifact is added, how would that impact other artifacts in terms 
of maintaining information consistent? \\
  8. Given that artifact A doesn't have any user OR is only used by role R, 
could the information in artifact A be merged into artifact B? \\
  \midrule
  \multicolumn{1}{c}{\emph{P2 - Artifact relationships}} \\
  9. How is the information in artifact A kept consistent with the information 
in artifact B, in the case C changes (C has two ''linked-to`` OR 
''used-to-create`` relationships to sibling artifacts A and B)? \\
  10. If inconsistencies between artifacts A and B arise, how does that impact 
the users of those artifacts and their work? \\
  11. What is the purpose of artifacts that are not related to any other 
artifact? \\
  12. Do the creators of artifact A deliver timely, i.e. can the 
information actually be accessed in ST when needed? \\
  \midrule
  \multicolumn{0}{c}{\emph{P5 - Artifact and role changes}} \\
  13. Does inconsistency of information among artifacts affect the work in: RE, 
ST, the interface between both? \\
  14. In case requirements change, by whom/how/when are these changes 
propagated to linked artifacts? \\
  15. How does staff turnover affect the quality of requirements and derivative 
artifacts? \\
  \midrule
  \multicolumn{1}{c}{\emph{P6 - Artifact and role scope}} \\
  16. Would involvement of RE/ST in creating artifact A improve the alignment?\\
  17. How is consistency between input from non-RE/ST artifacts and RE/ST 
artifacts maintained over time? \\
 \bottomrule
 \end{tabular}
\end{table}

\subsubsection*{Effort and Best Practices}
The budgeted effort for this step is 8-12 person-hours. Some effort 
(40\%) should be spent to represent the artifact map graphically. Representing 
the artifact map in a clear, easy to understand graph is crucial for the 
assessment workshop since the participants do not have much time to familiarize 
with complicated notation. In the cases studies, we used an all-purpose 
diagramming tool~\citep{yworks_yed_2014}, however have since then developed a 
dedicated tool\footnote{A prototype is available at 
http://lmsteiner.com/restbench} that supports the 
analyst in creating the artifact map.  The remainder of the effort should be 
spent on preparing a list of questions for the assessment workshop, using the 
prepared artifact map and the audio recordings of the interviews. Note that the 
audio recordings are not meant to be transcribed, but serve as source for 
details that were not captured yet in the map and to verify the created map. 
The list of questions should be separated into a clarifying part, addressing 
potential misunderstandings from the interviews, and an analysis part, seeded 
by the questions from Table~\ref{tab:questions}.

\subsubsection*{Example - Step 3}
Figure~\ref{fig:casee} shows the artifact map elicited at Company E. Note that 
for reasons of presentation clarity, we omit details of the complete artifact
map and show only the elements important for the analysis, discussed during the 
assessment workshop, in panels A and B (further discussed in 
Section~\ref{sec:collab}). The color coding (orange for RE, blue for ST and 
green for artifacts mentioned by both perspectives) illustrates the degree of 
agreement on the creation and use of artifacts. The commonly identified 
artifacts represent thereby the interface between RE and ST. Inconsistencies 
emerged from the data collection (and partially visible in the map) were picked 
up as entry points for analysis during the assessment workshop, for example:
\begin{itemize}
 \item Even though both RE and ST agree on Solution Definition and Development 
Requirements Specification (see Figure~\ref{fig:casee} and panel A therein), 
RE states that there is an explicit mapping between individual requirements and 
solution descriptions while ST claims there is no linking at all.
 \item The Test Strategy artifact was mentioned by the ST at Company E, but not 
by the ST at the supplier.
 \item A disagreement between RE of Company E and ST of the supplier emerged on 
the change frequency and time of the Development Requirements Specification.
\end{itemize}

These divergences between the RE and ST perspective on the artifact map are a 
result of the seeding questions P0 in Table~\ref{tab:questions} and are 
complemented with further questions originating from this seeding set, as shown 
in the next step.	

\subsection{Step 4 -- Assessment workshop}~\label{sec:collab}
The REST-bench assessment workshop is led by the analyst, introducing to all 
interviewees the process of the collaborative issue identification:
\begin{enumerate}
 \item Presentation and explanation of the artifact map 
 \item Error identification by interviewees
 \item Clarification questions by analyst
 \item Collaborative map analysis
 \item Summary and wrap up
\end{enumerate}

The analyst brings two printouts of the artifact map to the workshop, one for 
himself to note any corrections and one for the participants, explaining the 
notation and content of the artifact map. He should also point out immediately 
that the information in the map stems 
from the interviews, setting the context to the particular project that was 
selected for the data elicitation. The interviewees should spend 5-10 minutes 
with the map and try to identify any unclear artifacts, roles or incorrect 
relationships between artifacts. The analyst clarifies any 
uncertainties that emerged in the map construction\footnote{This is why map 
construction and collaborative issue identification in 
Figure~\ref{fig:restbench} are shown as iterative steps.}. The main task on 
which most of the allocated time should be spend on is to answer the questions 
prepared by the analyst, based on Table~\ref{tab:questions}. 

As for the initial data elicitation, we recommend to audio record the 
assessment workshop, while the analyst should take notes of the major 
identified issues. An issue might be related to the process (or not followed 
process), structure, content or distribution of artifacts, coordination between 
roles using artifacts, documentation infrastructure or lack thereof. It is 
important that all participants agree on an issue and the implications of it. 
The analyst should elicit evidence of these implications, e.g. in form of 
events during the project under discussion, from the workshop participants.

\subsubsection*{Effort and Best Practices}
The budgeted effort for this step is 6-10 person-hours. A realistic time-frame 
for the REST-bench assessment workshop execution is 2 hours. 

Even though the analyst initiates and drives the assessment workshop with 
prepared analysis points and questions, improvement opportunities are most 
likely to be acted upon when they emerge from the company employees. Therefore, 
the main goal of the analyst is to provoke an exchange between RE and ST, 
supported by the artifact map as a mean of communication, avoiding however 
conflicts and steer the discussion into a constructive direction. During the 
assessment workshop it is also important to maintain the context of the studied 
project, even though extrapolation to other projects is possible in order to 
understand the impact of an identified improvement opportunity. 

\subsubsection*{Example - Step 4}
\begin{figure}[t]
 \centering
 \includegraphics[width=\columnwidth]{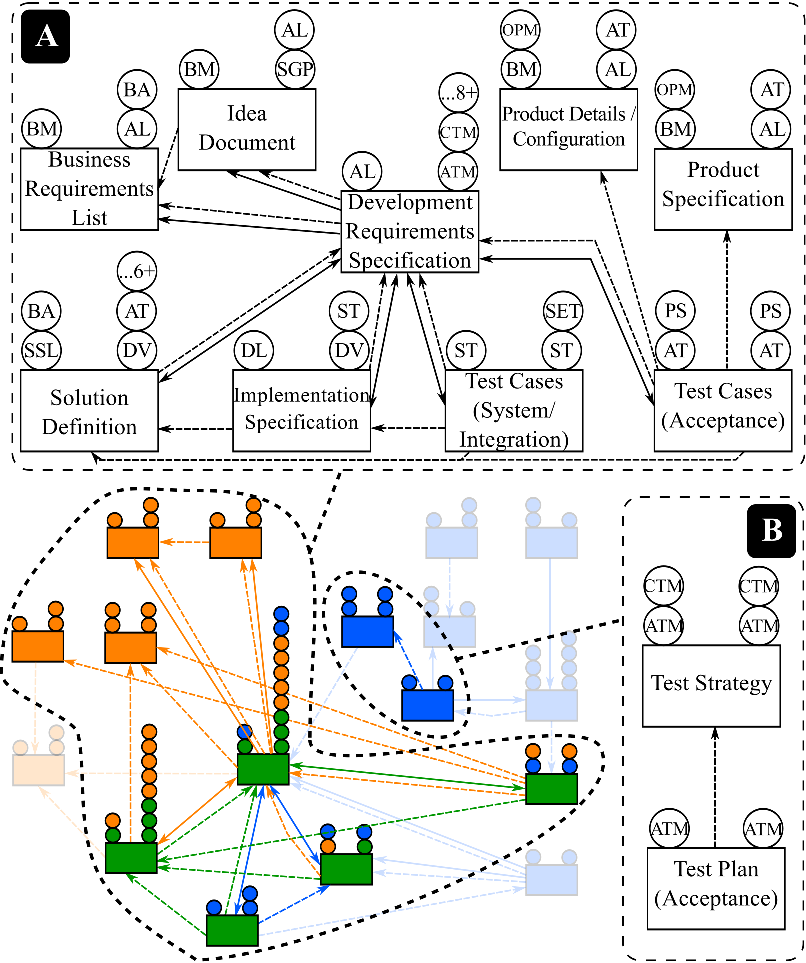}
 \caption{Artifact map of Case E}\label{fig:casee}
\end{figure}
Looking at Panel A in Figure~\ref{fig:casee}, the analyst observed that the 
Development Requirements Specification (i.e. the requirements 
therein) is mapped to the central artifacts in acceptance testing, system and 
integration testing, and development (Implementation Specification and Solution 
Definition). This mapping allows the company to assess the coverage of 
requirements both in terms of testing and implementation, enabling the 
monitoring of both development and testing progress. The analyst argued that a 
mapping between individual business requirements, e.g. from the 
Business Requirements List or the Idea Document, and the Development 
Requirements Specification would be important in order to know which business 
requirement corresponds to a particular development requirement. Indeed, RE 
confirmed during the workshop that it is not always clear to them which 
business requirements are already in production.

RE pointed out that such a mapping is difficult to achieve, since the 
Business Requirements List is not a standardized document and does also contain 
information not relevant for the requirements analysis. RE stated that the 
naming of use cases (part of the Development Requirements Specification) should 
describe the corresponding business requirement, creating an implicit mapping. 
According to RE, this is however not trivial and requires a lot of thought, as 
the use cases must illustrate the business process such that the use cases can 
be used ``as-is'' documentation for other projects. The analyst concluded that 
different roles have different requirements on the use cases and their names, 
making them not necessarily ideal for determining coverage of business 
requirements. Therefore, it might be better to pursue an explicit mapping, 
requiring however a formalization of the business requirements.

RE stated that it is sometimes difficult to get the Development Requirements 
Specification reviewed by the customer (internally, this is a Business Manager 
who requests a set of features). It is the customers' responsibility to look 
into the Development Requirements Specification and understand what they have 
agreed upon and verify the coverage of the business requirements. RE stated 
that even though it worked well in this project, it is still challenging in 
general to get customers to read the Development Requirements Specification and 
confirm that the business requirements are covered. The analyst concluded that 
an explicit mapping between business and development requirements would render 
the coverage analysis more effective.

Looking at Panel A in Figure~\ref{fig:casee}, the analyst observed a second 
issue: on the acceptance level, test cases are created by using information 
from the Product Specification, Product Details/Configuration, Development
Requirements Specification and the Solution Definition. Business requirements, 
represented in the Business Requirements List and the Idea Document are not 
used at all. Given the lack of mapping between business requirements and 
solution requirements, resulting in a potentially unknown business requirements 
coverage, the analyst concluded that using business requirements as input for 
creating acceptance test cases would be a good idea.

This issue was controversially discussed during the assessment workshop. On one 
hand, ST stated that it should not be required to use the business requirements 
directly as input, since all business requirements selected for the project 
should be represented in the Development Requirements Specification and the use 
cases therein. On the other hand, RE and the analyst argued that the 
Development Requirements Specification and the Solution Definition in 
particular are documents of the solution space, tending to describe the 
designed solution rather than the problem. The purpose of 
acceptance tests is to verify that the problem has been addressed by the 
solution whereas the purpose of system/integration tests is to verify that the 
solution implements the requirements correctly~\citep{bourque_guide_2014}.
The analyst concluded that a potential consequence of broadening the purpose of 
acceptance tests is that verification is duplicated, e.g. covering aspects that 
have been already covered in system tests. Indeed, RE confirmed that this can 
happen whereas ST stated that this is not the case in general.

Less controversial was the observation made by the analyst on Panel B in 
Figure~\ref{fig:casee} that there is no RE role involved in defining the Test 
Strategy or the Test Plan. Both RE and ST agreed that input from the RE 
perspective would be beneficial. A review of these artifacts by RE would 
validate that the tested functionality is the actually required functionality.

\subsection{Step 5 -- Report recommendations}\label{sec:rb-recommendations}
The purpose of this step is to summarize the findings from the workshop and to 
communicate them to a wider audience. For that reason, the report should be 
concise while still providing enough context in order to be useful for 
employees that did not participate in the assessment or in the studied project.

\subsubsection*{Effort and Best Practices}
The budgeted effort for this step is 8-12 person-hours. 
We recommend to provide a short introduction of the assessments purpose and 
scope. Then, the artifact map, based on the data collected in the interviews, 
should be presented, highlighting inconsistencies and analysis points relevant 
for the assessment workshop. The workshop summary should answer these analysis 
points, summarize the identified improvement opportunities, pointing to 
evidence in the updated artifact map, and eventually conclude with a set of 
recommendations. The report should be send to the study participants for review 
before it is communicated to a wider audience, allowing for corrections by the 
participants.

\subsubsection*{Example - Step 5}
In summary, based on the observations made during the collaborative workshop 
(see the example in Step 4), three areas for improvement were identified:
\begin{itemize}
 \item Business requirements gap: one of the Business Managers tasks is the 
 review of the Development Requirements Specification in order to verify 
 coverage of business requirements. The effectiveness of this review depends on 
 the skill/capability of the customer in understanding the Development 
 Requirements Specification. One improvement would be to strengthen these 
 skills by training. On the other hand, creating an explicit mapping from 
 Development Requirements Specification to business requirements would remove 
 the need for review. However, such a solution would require a formalization of 
 the Business Requirements List.
 \item Acceptance and System/Integration Test alignment: there are indications
that these test-levels have at least a certain overlap in what they actually
verify. As long as this overlap is intentionally created, the additional effort
can be controlled, otherwise this is an area where test efficiency can be 
improved.
 \item Involvement of the RE perspective, e.g. involving RE in Test 
Strategy and Test Plan reviews would help in validating that the development 
requirements are tested correctly.
\end{itemize}

\section{Case studies}\label{sec:casestudies}
In this section we present the four remaining cases where we applied 
REST-bench, chronologically such that improvements in the method follow 
logically from the described cases. Each subsection introduces first the case 
company, summarizes the assessment results and discusses the impact the lessons 
learned had on REST-bench.

\subsection{Company A}\label{sec:casea}
We selected a system manager with 12 years experience in his current role as RE 
representative. For the ST representative we chose a verification engineer with 
14 years experience. The project in which the two engineers collaborated 
completed in autumn 2011, while we performed the assessment in June 2012. The 
project staff consisted of 150 engineers, split up into seven teams. The system 
requirements consisted of approximately 350 user stories, whereas the system 
test cases amounted to 700, of which 550 were automated.

\subsubsection{Assessment results}
\begin{figure}[t]
 \centering
 \includegraphics[width=\columnwidth]{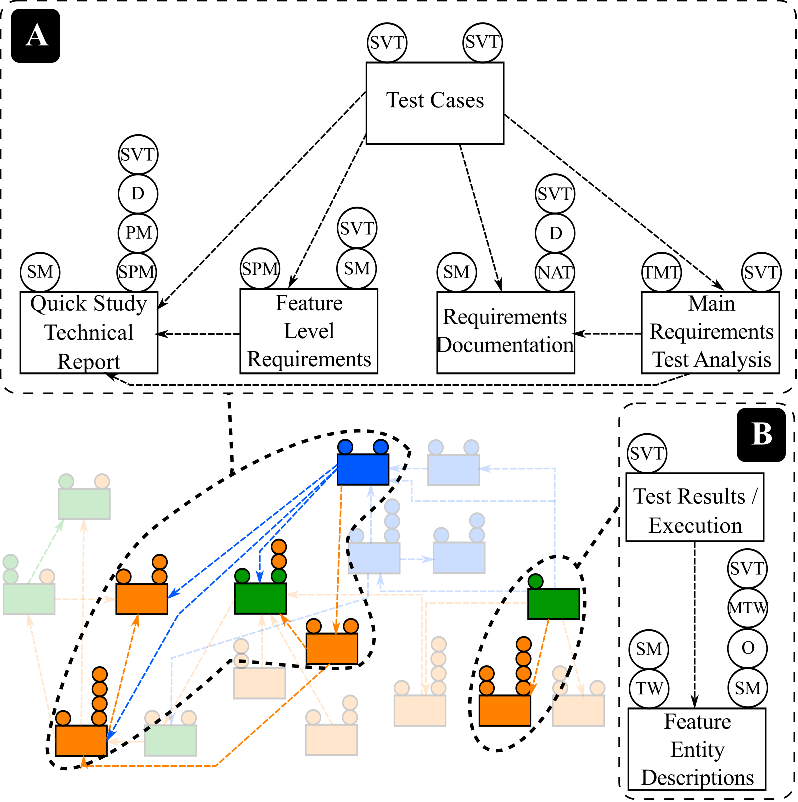}
 \caption{Artifact map of Case A}\label{fig:casea}
\end{figure}
Figure~\ref{fig:casea} shows the artifact map used during the REST-bench 
assessment workshop. For clarity, we highlight only those details of the map 
(Panel A and B) relevant for the identified improvement opportunities. 

Panel A in Figure~\ref{fig:casea} illustrates that the Software Verification 
Team (SVT) uses four artifacts as input to create the Test Cases. The 
Requirements Documentation is the main source, complemented by the other 
artifacts. During the assessment workshop the analyst questioned whether 
inconsistencies in these documents, caused for instance by requirements changes 
during the design or implementation, affect ST. According to RE, changes in 
user stories (part of the Requirements Documentation) are propagated to the 
SVT, which is invited to the presentations where changes are discussed. In the 
context of this particular project, RE stated that parts of the solution where 
redesigned late in the project, when the development already had started. This 
was confirmed by ST, asserting that there were inconsistencies between Test 
Cases and Requirements Documentation due to a too early test analysis. 

The Requirements Documentation artifact was elicited from both interviewees and 
represents therefore an important interface between RE and ST. According to ST, 
there is however no linking between Requirements Documentation and Test Cases 
at the Integration Test level. This is due do the difficulty to keep these
links up-to-date; early attempts with spreadsheets failed and importing the 
required information into the test management tool is labor intensive. 
According to RE, this lack of linking may lead to lower test coverage of the
requirements. ST stated that the lack of traceability is to some extent 
compensated by the Technical Manager for Test (TMT) whose responsibility it is 
to define the test scope. The analyst concluded that the TMT acts thereby as a 
link between RE and ST. However, testers need to pull information from TMT, 
rendering the spread of knowledge on requirements changes a matter of 
individual initiative, that is, person dependent. 

Panel B in Figure~\ref{fig:casea} shows that Feature Entity Descriptions, 
describing the system functionality on a compound level, are used by the SVT to 
interpret the Test Results and Execution. However, RE stated that Feature 
Entity Descriptions are written late in the project, by external consultants, 
describing how the system is actually implemented. RE stated that this is a 
local sub-optimization that saves effort in the RE department, affects however 
the work of ST. The analyst concluded that, to render Feature Entity 
Descriptions useful to ST, they should be created and maintained during the 
project as the implementation stabilizes. To summarize, the following 
improvement opportunities were identified:
\begin{itemize}
	\item Test Cases are created by using different sources of information 
	which are potentially inconsistent.
	\item Requirements changes, tracked by a dedicated management role (TMT), 
	should be propagated to the responsible test engineers.
	\item Feature Entity Descriptions should be written earlier such that they 
	are of more use to ST.
\end{itemize}

\subsubsection{Impact on REST-bench}
We intended to collect and analyze both artifacts (tangible information)
and events when information is exchanged informally, e.g. in ad-hoc meetings, 
email and on-line conversations, or phone calls. However, eliciting each 
individual instance of informal communication seemed unrealistic for the 
targeted effort budget (6-12 person-hours) for data elicitation.
The positive results of this first assessment, i.e. the identification of 
issues, supported our intuition that we already collect enough data that can be 
efficiently analyzed in the assessment workshop and lead to concrete 
improvement suggestions.

Regarding the relationships between artifacts, we aimed at producing
rich analysis points, identifying \emph{what} information is used by 
\emph{whom} to create \emph{which} artifact. Therefore we focused to elicit 
information regarding ``used-to-create'' relationships. However, using an 
artifact to create another does not necessarily mean that the two artifacts are 
linked, rendering the REST taxonomy 
heuristics~\citep{unterkalmsteiner_taxonomy_2014} more difficult to apply. The 
workshop in Company A revealed that links between artifacts, or lack thereof 
(in that case, Requirements Documentation and Test Cases at the Integration 
Test level), would also be useful to analyze. Therefore we decided to also 
elicit ``linked-to'' relationships between artifacts and present them as solid 
lines in the map. Note that in the following two assessments, at Company B and 
C, we focused on eliciting only ``linked-to'' relationships (partially to 
maintain the elicitation time budget, partially to keep the artifact map easy 
to interpret). However, we realized that ``used-to-create'' relationships 
between artifacts are useful for the analysis (see 
Section~\ref{sec:companyCimpact}), and introduced them again in the assessments 
of Company D and E.

Finally, we relied exclusively on note taking during the interviews. Even 
though this sped up the artifact map creation process, much time during the 
assessment workshop was spent on correcting the map. Hence we decided to audio 
record future interviews and workshops, given that participants gave consent.

\subsection{Company B}\label{sec:caseb}
In Company B we could not identify a single project, since Scrum teams (one 
business analysis and two development/test teams) work continuously on a 
product, releasing a new version every two months. We chose therefore the last 
release-cycle as time-frame for the assessment and selected a requirements 
engineer and a test lead as interviewees. The total staff consisted of 
20 engineers while the assessment took place in March 2013. 

\subsubsection{Assessment results}
\begin{figure}[t]
 \centering
 \includegraphics[width=\columnwidth]{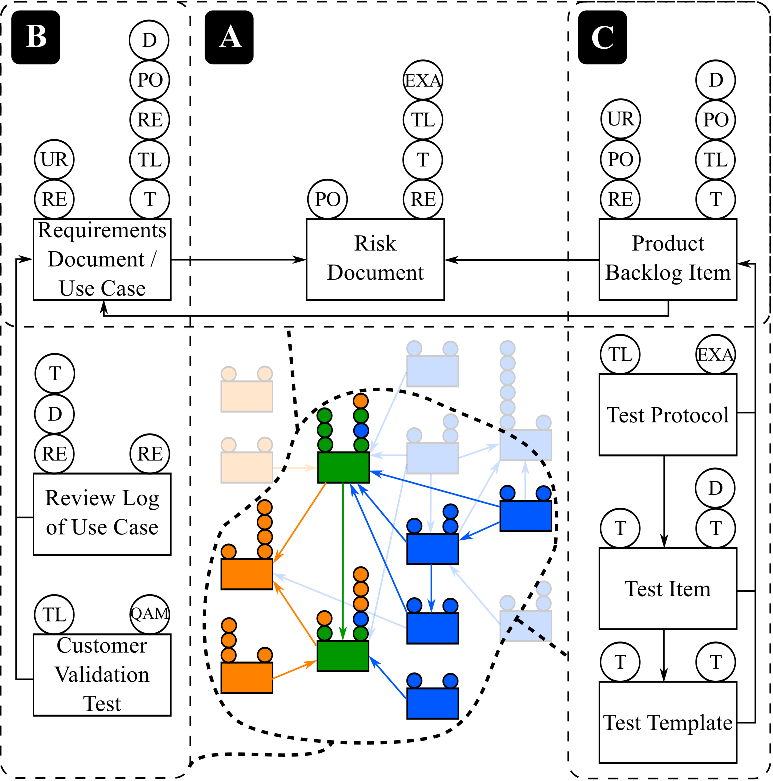}
 \caption{Artifact map of Case B}\label{fig:caseb}
\end{figure}
RE stated that regulatory requirements enforce that the development process  
provides traceability, illustrating that safety risks are considered 
in the requirements, and that the developed services or products comply to 
these requirements (verification).
However, according to RE, maintaining traceability over time is complicated by 
storing information in different formats. While all artifacts are stored in a 
content management system providing revision control, creating and maintaining 
links between artifacts is not supported by the system. ST stated that this 
complicates impact analysis (even though traces between artifacts exist, they 
need to be found manually as the content management system does not handle 
traces), e.g. when a requirement changes and the corresponding Test Items and 
Test Templates need to be updated or rerun. The analyst concluded that this 
``infrastructure gap'' may lead to an increased effort in impact analysis and 
to potential inconsistencies in test artifacts. 
Another related issue is the redundancy of manual traceability links. For 
example, looking at Panel A in Figure~\ref{fig:caseb}, both Product Backlog 
Item and Use Case refer to a particular risk. RE stated that these links are 
maintained manually, causing additional effort to check consistency. ST stated 
that inconsistencies in these manual links may be caught (early) during the 
review of the use case or (late) when the sprint is approved. RE and ST agreed 
that it would be sufficient to link the Risk Document to Product Backlog Items.

While traceability is generally established with unique references, ST stated 
that Customer Validation Tests are linked implicitly to Use Cases by naming 
conventions (Panel B in Figure~\ref{fig:caseb}). These naming conventions, 
created by the Test Lead, are however not enforced or documented.  The analyst 
concluded that staff turnover may therefore easily break this link. 
Furthermore, as soon as the complexity of the validation tests 
rises, a more sophisticated tracing mechanism may be required in order to keep 
track whether and how the Use Case is covered by the executed scenarios in the 
Customer Validation Tests.

Looking at Panel B in Figure~\ref{fig:caseb}, the analyst observed that ST 
roles are involved in reviewing documents created by RE. There is however no 
such involvement of RE roles in reviewing documents created by ST (Panel C). RE 
stated that they occasionally, when a complex function needs to be verified, 
provide feedback to ST on particular test cases. Such ad-hoc reviews do however 
not verify the test scope. The analyst concluded that this could be achieved by 
having a review of the Test Template and the test-cases therein (early, before 
the tests are actually executed), or of the Test Protocol (late, when the test 
results are available). To summarize, the following improvement opportunities 
were identified:
\begin{itemize}
	\item Manual maintenance of traceability links leads to increased effort in 
	impact analysis. Redundant, manually maintained, traceability links may 
	lead to inconsistencies requiring rework.
	\item Customer Validation Tests are not explicitly mapped to Use Cases.
	\item RE is little involved in reviewing document created by ST.
\end{itemize}

\subsection{Company C}\label{sec:casec}
\begin{figure}[t]
 \centering
 \includegraphics[width=\columnwidth]{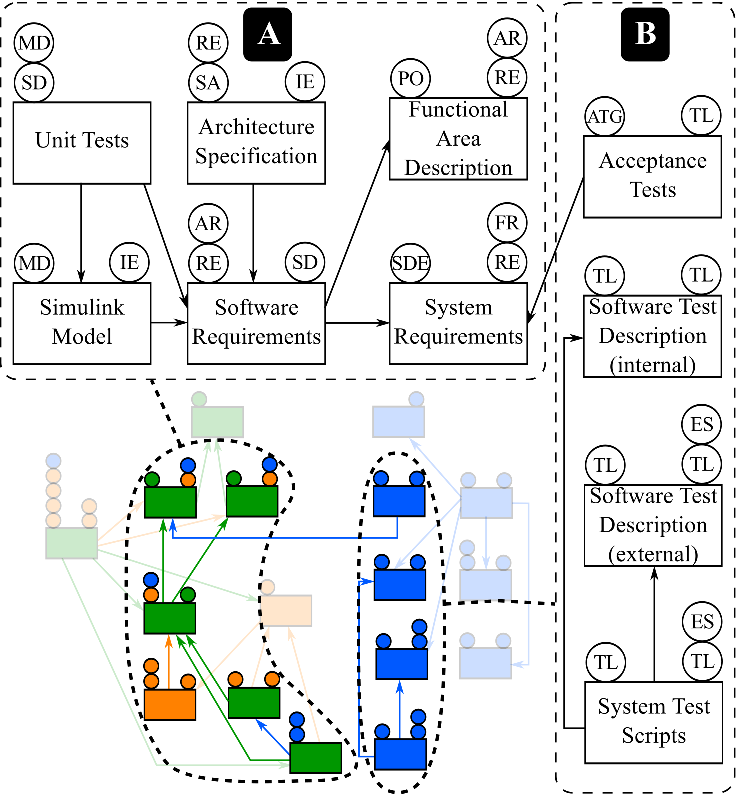}
 \caption{Artifact map of Case C}\label{fig:casec}
\end{figure}
We selected a requirements engineer/software architect with 3 years experience 
in his current role as RE representative. For the ST representative we chose a 
test lead with 1.5 years experience. The assessment took place in March 2013, 
approximately at halftime of the expected three year project duration. The 
selected project had a staffing of 9 engineers, operating in a Scrum team. The 
overall development process in Company C is plan-driven, posing challenges to 
the project that iterates in two week sprints. The project was responsible for 
approximately 300 system requirements.

\subsubsection{Assessment results}\label{sec:c_results}
An aspect that is not visible in the artifact map is the project life-cycle and 
how it influences the artifacts that are created and used. This project was
estimated to last three years in total. This investigation took place 
approximately 1.5 years into the project, at a time when the high-level 
requirements (System requirements and Functional Area Descriptions, see Panel 
A in Figure~\ref{fig:casec}) were stabilized and soon to be
frozen. Due to the volatility of the requirements in the initiation phase of 
the project, the development team focused on establishing the feasibility of 
the product, adapting continuously on changing requirements. This led to 
postponing the formalization of documentation and links, since this process was 
time consuming and there was a resistance to start it before there was an 
indication that the system was actually working. Therefore, even if Panel A in 
Figure~\ref{fig:casec} suggests that traceability from high-level requirements 
down to unit-tests is given, this was only a goal that has not yet been 
achieved at this stage of the project. In particular, the links from 
the Software Requirements to System Requirements and Functional Area 
Descriptions were in the process of being re-engineered and implemented in the 
requirements specification tool (SpecTool). 

Looking at Panel B in Figure~\ref{fig:casec}, the analyst observed that there 
is a gap between the integration tests (Software Test Description artifact) and 
the requirements they are intended to verify (System Requirements and 
Functional Area Description). Software Requirements had, according to RE, a 
1-to-1 mapping to unit-tests. On the other hand, System Test Descriptions, 
which specify how and what was tested in integration, were not linked to any 
requirements documentation. Both RE and ST agreed that linking the System Test 
Descriptions to System Requirements and Functional Area Descriptions would 
provide benefits:
\begin{itemize}
 \item ability to perform requirements coverage analysis
 \item change impact analysis, leading to reduced test-time
 \item allow reviews in which both RE and ST participate and validate the
effectiveness of integration tests
\end{itemize}

In this project, according to RE, the software team started to integrate System 
Requirements and Functional Area Descriptions in SpecTool, allowing them to 
link their information to Software Requirements. Although this was an 
improvement (as manual linking to word documents was not necessary anymore), 
linking the requirements to the integration tests was still a challenge, since 
test artifacts were stored in a separated infrastructure, requiring a manual 
linking and maintenance process.
According to RE, the link implementation in SpecTool by itself may turn out to 
be problematic. Even minor changes in parent documents, such as correction of 
spelling mistakes or improving descriptions, would trigger a change event,
leading to child documents to be flagged as outdated. There was no possibility 
in SpecTool to qualify the level of a change. That may be a risk, leading to a 
lower quality of documentation. To summarize, the following improvement 
opportunities were identified:
\begin{itemize}
	\item Time consuming re-engineering of Software Requirements and linking to 
	other artifacts.
	\item Lack of traceability between integration tests and requirements they 
	are intended to verify.
\end{itemize}

\subsubsection{Impact on REST-bench}\label{sec:companyCimpact}
As we assessed Company B and C in parallel, we summarize here the lessons 
learned from both REST-bench applications. As a major change, we reintroduced 
the elicitation of ``used-to-create'' relationships between artifacts, adding 
how information from one artifact is used to create another artifact. This 
differs from the ``linked-to'' semantics and is useful as it provides a 
dynamic aspect to artifact relationships that is lost when considering only 
links between artifacts. For example, consistency between artifacts is much 
more difficult to achieve and maintain when there exist 
``used-to-create'' but no ``linked-to'' relationships between artifacts. 

Nevertheless, the assessment in Company C illustrated also one of the 
weaknesses of a static artifact map. At different points in time, the 
particular use of an artifact may change in a project. This dynamism is not 
visible in an artifact map, but can be captured during data collection and then 
considered while discussing the map at the assessment workshop.

\subsection{Company D}\label{sec:cased}
We selected a requirements engineer and a system designer, with 10 and 5 years 
experience respectively, as RE representatives. We chose a test lead with 5 
years experience as representative for the ST perspective. The selected project 
had a four year duration, involving in peak times about 1000 hard-and software 
engineers. During the assessment on April 2013, the project was in the closing 
phase. The project handled approximately 2000 system requirements and 500 
functional test cases.

\subsubsection{Assessment results}
\begin{figure}[t]
	\centering
	\includegraphics[width=0.9\columnwidth]{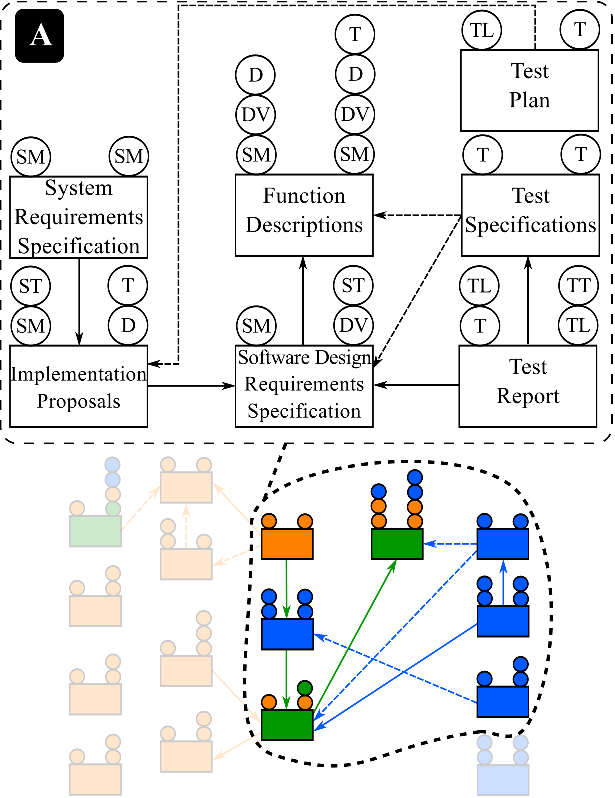}
	\caption{Artifact map of Case D}\label{fig:cased}
\end{figure}
RE stated that Implementation Proposals~\citep{fricker_handshaking_2010} were 
typically written as ``change this existing functionality like that'', in 
collaboration by RE and ST roles, and were used by system designers to 
architect the software and by ST to develop the test plan (see Panel A in 
Figure~\ref{fig:cased}). Function Descriptions, on the other hand, were written 
when the software has been implemented, providing a complete description of a 
function, and were used by ST to write test specifications.
ST reported that the information in the Implementation Proposals and the 
respective Function Descriptions may be inconsistent, leading to situations 
where they attempted to test functionality that is not supported by the 
delivered software. Inconsistencies stemmed from the fact that Function 
Descriptions were created late in the project forcing ST to use the 
Implementation Proposal to write test-specifications, and that no role has been 
assigned the responsibility to check and maintain consistency between 
Implementation Proposals and Function Descriptions.

Due to the amount of Implementation Proposals and Functional Descriptions per 
project, maintaining consistency among them is a challenge and may not even be 
possible with reasonable effort. One possible improvement, mentioned by RE, 
would be to implement a mechanism that alerts the reader of potential 
inconsistencies. Furthermore, the responsibility (who, why) and the procedure 
(what, when) of updating/creating Function Descriptions should be defined.
The Implementation Proposal was written by both RE and ST, whereby ST was
responsible for the test specific parts. RE stated that the distance in time in 
which different parts in the Implementation Proposals were written has been 
very long (up to one year): while RE was working on an Implementation Proposal, 
ST might be heavily involved in testing, leading to the situation that ST
was actually not collaborating with RE in the specification. Then, when ST was
ready to specify the testing part, the document may have been out-of-date as it 
was not updated by RE during the project. RE proposed an improvement to the 
process, that is, to ensure the commitment of both RE and ST to collaborate at 
the same time on the Implementation Proposal specification. 

ST explained that in their work process, they mapped the requirements (from the
Software Design Requirements Specification) to the test cases they execute. 
In the Test Report they documented test results and listed which requirements 
were covered by the executed test cases (see Panel A in 
Figure~\ref{fig:cased}). 
There was however, according to RE, no feedback loop from ST to RE, leading to 
the inability to improve low quality requirements when they were marked as 
untestable in the Test Report. Furthermore, test-coverage could not be 
efficiently measured. Both RE and ST agreed that a requirement to test-case 
mapping would have the following benefits:
\begin{itemize}
 \item Regression tests could be efficiently defined (assuming that Software 
Design Requirements Specification is traced to requirements at higher 
abstraction levels).
 \item Change requests involving the addition or removal of requirements would
be visible on the ST side, i.e. it would be transparent whether new test cases 
are needed, or old ones can be removed.
 \item Statements of verification could be generated without manual effort.
\end{itemize}

Looking at the artifact map in Figure~\ref{fig:cased}, the analyst observed 
that ST was, except in the definition of Implementation Proposals where the 
benefit is immediate, not involved in RE activities. RE stated that previous 
efforts to include ST had failed, mostly due to scheduling issues. With the 
initiative to improve the quality of the Software Design Requirements 
Specification, new opportunities for ST involvement could arise. For example, 
ST could provide useful input on the requirements quality by looking at their 
Test Reports. Not testable requirements identified in these reports can either 
be improved or even removed from the database.
A second opportunity for ST involvement arises as soon as the requirements to
test case mapping is implemented. New requirements need to be reviewed and
mapped to either existing test cases or the decision needs to be taken to create
new test cases. ST could be assigned this responsibility.
Then, with the requirements to test case mapping in place, ST can provide
feedback, during and after the testing phase, to RE on the quality of
requirements, preventing the decay of requirements quality. To summarize, the 
following improvement opportunities were identified:
\begin{itemize}
	\item Maintaining consistency between Improvement Proposals and Function 
	Descriptions in order to support ST in testing the correct functionality.
	\item Schedule collaboration between RE and ST such that work on common 
	artifacts leads to coordination rather than misunderstandings.
	\item Map test cases to requirements such that faults in requirements 
	identified by ST can be fed back to RE.
\end{itemize}

\subsubsection{Impact on REST-bench}
In Company D, we analyzed the lack of mapping between individual requirements 
and test cases. Even though requirements are linked in the Test Report to 
individual test cases, RE could not benefit from that information as no 
bi-directional links existed. This observation, together with the analysis 
of Company C (see Section~\ref{sec:c_results}), where we did not differentiate 
between uni-directional and bi-directional relationships, led us to the 
introduction of the ``mapped-to'' relationship. The semantics of the 
``mapped-to'' relationship allows us to describe links between artifacts that 
can be followed from either side (corresponding to backward and forward 
traceability). 

\section{Discussion}\label{sec:discussion}
Looking at the assessments performed with REST-bench, one commonality 
encountered in all five cases is the identification of improvement 
possibilities in the linking mechanisms between nodes of information. This 
means either to establish a link in the first place, or improve an existing 
link such that it is more resilient, accurate or reliable over time. In some 
cases, a solution would be to introduce explicit traces between artifacts, 
either by adopting a viable trace model~\citep{ramesh_toward_2001} or by 
automatically recovering trace links (see \cite{borg_recovering_2014} for an 
overview of techniques). In other cases, the involvement of test engineers in 
requirements engineering tasks would be beneficial, as observed by 
\cite{damian_empirical_2006}, \cite{uusitalo_linking_2008} and 
\cite{kukkanen_applying_2009}. Generally, the assessed companies
were well aware that information either from RE or ST is not used to its
full potential for decision support. Consequentially, the study participants
valued REST-bench's ability to identify gaps in the coordination between
RE and ST. Similar observations, focused on gaps between RE and downstream 
development affecting test coverage and scoping were made by 
\cite{bjarnason_requirements_2011} in the context of large-scale software 
development. However, since the assessments did not include solution 
development, monitoring and evaluation of the implemented changes, we do not 
have resilient results that would allow us to provide general recommendations 
on how to improve REST alignment. In the remainder of this section we answer 
the research questions stated in Section~\ref{sec:rm}.
\begin{table*}[t]
 \footnotesize
 \caption{Post assessment questionnaire}\label{tab:questionaire}
 \begin{threeparttable}
 \begin{tabular}{lp{0.75\textwidth}l}\toprule
 \multicolumn{3}{c}{Open ended questions} \\
 \midrule
 Q1 & \multicolumn{2}{p{0.9\textwidth}}{Did the elicitation (interview 
session), the artifact map or the collaborative workshop reveal something new, 
you did not know before, w.r.t. activities, responsibilities or artifacts in: 
requirements engineering, software testing, the interplay between both?} \\
 Q2 & \multicolumn{2}{p{0.9\textwidth}}{Given the resources you have invested, 
would you consider to use this method as a complement to project post-mortems 
to improve the coordination between RE and ST? Please motivate, why 
yes/no/maybe.} 
\\
 Q3 & \multicolumn{2}{p{0.9\textwidth}}{Did you discuss the coordination 
between RE and ST in your organization already before this assessment? If yes, 
which specific issue(s) have been discussed and why?} \\
 Q4 & \multicolumn{2}{p{0.9\textwidth}}{In general, to elicit more relevant and 
precise information, would you suggest to include another role (e.g. Software 
Architect, Developer) in the assessment? Please motivate, why yes/no/maybe.} \\
 \midrule
 \multicolumn{2}{c}{5 point Likert scale with possibility to provide 
explanatory feedback} & Results\tnote{3}\\
 \midrule
 Q5 & The assessment (including elicitation and workshop) is an 
\emph{efficient}\tnote{1}\, mean to identify issues in relation to the 
coordination between Requirement Engineering and Software Test. &
\tablepicture{data/q5.dat}\\
 Q6 & The assessment (including elicitation and workshop) is an 
\emph{effective}\tnote{2}\, mean to identify issues in relation to the 
coordination between Requirement Engineering and Software Test. & 
\tablepicture{data/q6.dat} \\
 Q7 & The artifact map and the discussion in the workshop increased my 
understanding of the coordination between RE and ST in the studied project. & 
\tablepicture{data/q7.dat}\\
 Q8 & The artifact map and the discussion in the workshop increased my 
awareness for potential waste (e.g. unused documentation). & 
\tablepicture{data/q8.dat}\\
 Q9 & The artifact map and the discussion in the workshop increased my 
awareness for potential gaps in the coordination between RE and ST. & 
\tablepicture{data/q9.dat}\\
 Q10 & Interactions (scheduled meetings, but also casual encounters over 
coffee, in the corridor or at the office door) are equal or even more important 
as written documentation for the successful execution of the project. & 
\tablepicture{data/q10.dat}\\
 \bottomrule
 \end{tabular}
 \begin{tablenotes}
  \item[1] Efficiency in general describes the extent to which time or effort 
is well used for the intended task or purpose. 
  \item[2] Effectiveness is the capability of producing a desired result. When 
something is deemed effective, it means it has an intended or expected outcome.
  \item[3] The bars represent the 10 responses from the study participants on a 
  Likert scale (strongly agree, agree, neutral, disagree, strongly 
  disagree)
 \end{tablenotes}
 \end{threeparttable}
\end{table*}

\subsection{RQ1: To what extent are the dyad structures from the 
REST taxonomy useful to elicit improvement opportunities?}

Table~\ref{tab:questions} illustrates the mapping between the seeding questions 
we used to prepare the assessment workshops and the dyad structure properties, 
summarized in Table~\ref{tab:dsp}, we identified in our earlier 
work~\citep{unterkalmsteiner_taxonomy_2014}. We established the mapping in 
Table~\ref{tab:questions} in order to understand which dyad structure 
properties are actually useful in generating analytical questions regarding the 
elicited artifact maps. While properties \emph{P1}, \emph{P2}, \emph{P5} and 
\emph{P6} were useful, we could not yet derive questions by using properties 
\emph{P3} and \emph{P4} (see Table~\ref{tab:questions} for a description of 
these properties). However, this does not preclude the possibility that in 
future assessments those properties might generate useful seeding questions. 

Looking at the overall focus of REST-bench, it seems straightforward why no 
questions regarding \emph{P3}, intermediate nodes, were generated: the data 
elicitation is geared towards artifacts that are used and created by 
requirements engineers and software testers (see Section~\ref{sec:elicitation}). 
Q4 in our post assessment questionnaire addressed this specific aspect, i.e. 
whether other roles (e.g. software architects, developers) should be included 
in the assessment. Nine out of ten participants suggested including other roles 
(six argued for software developers, one for a configuration manager, two for 
no specific role). While we decided not to include more roles, in order to keep 
the overall assessment effort low, one could argue that a developer 
participating in the assessment workshop alone would be beneficial as a user of 
requirements and creator of the system under test, thereby contributing to the 
identification of improvement potential.

The second dyad structure property for which we could not generate analytical 
questions is \emph{P4} - RE and ST node proportion. This property expresses the 
relative effort spent in creating and maintained RE and ST artifacts 
respectively. However, we did not find any evidence that this proportion 
could indicate an issue in REST alignment or serve as an analytical lever to 
identify improvement opportunities.

To gauge the overall relevance and usefulness of REST-bench, we asked questions 
Q1-Q3 (see Table~\ref{tab:questionaire}) to participants. Regarding the 
relevance of REST alignment (Q3), eight out of ten participants reported that 
they had in the past discussions on how to better coordinate requirements 
engineering and testing activities. As one participant put it, ``\emph{we 
have had some discussions in my team how to get the correct information when we 
need it from RE. But we have not discussed that with the RE organization 
yet.}'' All ten participants stated that they would use REST-bench as a 
complement to project postmortems (Q2), as ``\emph{it gives a complete picture 
of the artifacts, the relations between them and a good material for use in 
discussions about potential process improvements}'', and ``\emph{is a very good 
and easy way to ensure that all involved use and acknowledge each others 
documents and their purpose.}'' This is further corroborated by the results on 
Q5 and Q6, where we asked the participants whether REST-bench is efficient and 
effective (see Table~\ref{tab:questionaire}). One participant stated that 
``\emph{the method gives more truth on what is actually used. We already have 
processes for a lot of things but these are somewhat theoretical and not 
adapted per project.}'' On the other hand, another participant stated that 
``\emph{it is not really guaranteed that we take action just because we have 
identified the issues.}'' With respect to Q1, whether participants learned 
something new from the artifact map or the assessment workshop, 8 out of 10
participants indicated that they identified aspects in the way of working they 
were not aware of. This is further supported by the results to Q7 in 
Table~\ref{tab:questionaire}, even though a participant stated that ``\emph{I 
believe I already had a pretty good overview of the situation. Probably because 
we are a relatively small team where everybody works with a broad range of 
tasks}'', and some other stated that ``\emph{it confirmed my thoughts}.'' 
Another aspect we investigated is whether REST-bench increases the awareness of 
potential waste (Q8) or gaps (Q9) in the coordination between RE and ST. The 
results (see Table~\ref{tab:questionaire}) suggest that the potential is more 
geared towards identifying gaps than waste. REST-bench can be used to identify 
waste candidates~\citep{khurum_extending_2014}. However, to actually remove 
waste all project stakeholders are necessary. This corroborates the result from 
Q4 discussed earlier, including developers could improve the potential of 
REST-bench to eliminate waste. 

\subsection{RQ2: To what extent is REST-bench in Agile and plan-driven
environments useful?}
One of our concerns when we planned the validation of REST-bench was the 
methods' focus on using documentation as a proxy to determine REST alignment. 
Indeed, in Case A described in Section~\ref{sec:casea}, we planned to 
complement the data collection with recording instances of informal 
communication, however rejected the idea due to the involved effort and 
inaccuracy to do that manually. Therefore, we questioned whether we can apply 
REST-bench in Agile environments at all, and collect relevant data and produce 
useful results. Agile approaches are notorious for promoting as little 
documentation as 
possible~\citep{fowler_agile_2001,chau_knowledge_2003,cohen_introduction_2004, 
nerur_challenges_2005}. 

In order to understand whether there is a difference in usefulness of 
REST-bench, we analyzed the post assessment questionnaire by stratifying the 
answers into plan-driven and Agile groups, using the project characteristics 
illustrated in Table~\ref{tab:companies}. There were five 
participants in each group. The largest divergence can be observed in Q8 
(REST-bench increases the awareness for potential waste). The Agile group tends 
to disagree~(\picture{data/q8a.dat}), while the plan-driven group tends to 
agree~(\picture{data/q8t.dat}). On the other hand, both the 
Agile~(\picture{data/q9a.dat}) as well the 
plan-driven~(\picture{data/q9t.dat}) group agree that 
REST-bench increases the awareness of gaps in the coordination between RE and 
ST (Q9). Since REST-bench assesses coordination through the creation\slash use 
of documentation, this result from the Agile respondents might seem surprising. 
However, this conforms to the observations made in a survey among Agile 
practitioners where the majority reported that documentation is important or 
very important~\citep{stettina_necessary_2011}. Both groups show similar 
tendencies to the remaining questions:
\begin{itemize}
 \item Q5 - efficiency of REST-bench: Agile~(\picture{data/q5a.dat}) and 
plan-driven~(\picture{data/q5t.dat})
 \item Q6 - effectiveness of REST-bench: Agile~(\picture{data/q6a.dat}) and 
plan-driven~(\picture{data/q6t.dat})
 \item Q7 - increased understanding: Agile~(\picture{data/q7a.dat}) and 
plan-driven~(\picture{data/q7t.dat})
  \item Q10 - importance of informal interactions: 
Agile~(\picture{data/q10a.dat}) and plan-driven~(\picture{data/q10t.dat})
\end{itemize}

These results suggest that REST-bench is useful both in Agile and plan-driven 
environments. This is further corroborated by the actual case assessment 
results, where for both Agile and plan-driven projects relevant improvement 
opportunities were identified.

\subsection{To what extent is REST-bench usable?}
\begin{table}
	\footnotesize
	\caption{Usability assessment of REST-bench}\label{tab:usability}
	\begin{tabular}{p{2cm}llp{3.3cm}}\toprule
		Step & Cost & Effort (p/h) & Support within REST-bench \\
		\midrule
		Selection & 5\% & 2-4 & heuristics on participant profile \\
		Data elicitation & 15\% & 6-12 & defined procedure / templates \\
		Map construction & 40\% & 8-12 & seeding questions / mapping tool 
		prototype \\
		Collaborative issue identification & 30\% & 6-10 & defined procedure \\
		Report recommendations & 10\% & 8-12 & report structure \\
		\bottomrule
	\end{tabular}
\end{table}

We answer this question from the perspective of the researcher who 
applied the method. REST-bench has not been transferred to an industry 
partner so that ease of use from the perspective of a practitioner can not 
be validated yet. However, the results from the questionnaire indicate that 
REST-bench is, from a participants' perspective, efficient and effective (see 
Q5 and Q6 in Table~\ref{tab:questionaire}). 

We look at three aspects of usability, breaking it down to each step in 
REST-bench: required up-front investment, the effort to perform, and the 
support REST-bench provides in each step. With up-front investment we mean the 
one-time cost for the analyst to learn a particular step. In 
Table~\ref{tab:usability}, we provide the relative cost of each step. In total, 
we estimate the up-front cost to learn the REST-bench method to 8-12 hours.

The required estimated effort is shown in the third column in 
Table~\ref{tab:usability}, and further motivated in the respective sections 
where each step is explained 
(Section~\ref{sec:rb-selection}~-~\ref{sec:rb-recommendations}). We regard 
these estimates as upper limits (with the suggested number of participants), if 
the assessment is performed in regular intervals. 
A common detractor for conducting postmortems is lack of 
time~\citep{keegan_quantity_2001,glass_project_2002}. Therefore, we designed 
REST-bench to be efficient, while still providing value to the organization 
conducting the assessment. This efficiency is achieved by providing guidelines 
and heuristics within REST-bench (see fourth column in 
Table~\ref{tab:usability}). Furthermore, each of the steps is supported by a 
series of examples presented in this paper.

\section{Conclusions}\label{sec:conclusions}
The paper presents a method to identify improvement opportunities in the 
coordination between requirements engineering (RE) and software testing (ST). 
The method, REST-bench, is based on the premise that information, the way it 
is created, used and linked, is key for understanding how requirements and test 
engineers coordinate and how their collaboration can be improved. We used the 
information dyad and the emerged dyad structure properties from our earlier 
work~\citep{unterkalmsteiner_taxonomy_2014} to drive an assessment process that 
is designed to be lightweight in terms of resource use (30-50 person-hours per 
assessment). 

We validated REST-bench by applying it in five companies. In all assessments we 
could identify issues that were taken up by the involved companies to trigger 
improvements. With respect to RQ1, whether the dyad structure properties from 
the REST taxonomy were useful, we found that all but two properties generated 
seeding questions that can be used to identify issues and elicit improvement 
opportunities. Furthermore, the feedback gathered from the assessment 
participants supports our conclusion that REST-bench is an effective method to 
identify gaps, and to a lesser degree waste, in the artifacts that support RE 
and ST coordination. With respect to RQ2, applying REST-bench in Agile and 
plan-driven contexts, we conclude that the proposed method is useful in both 
while it is more likely to identify waste in plan-driven contexts. Furthermore, 
projects with very small teams requiring little coordination will likely not 
benefit from a REST-bench assessment. Answering RQ3, we found that REST-bench 
is usable from the perspective of the analyst who conducted the assessments. 

To corroborate these results we plan to further validate the usefulness and 
usability of REST-bench by training practitioners in the autonomous use of the 
method, supported by efficient tools for data collection and artifact mapping.
.

\section*{Acknowledgments}
We would like to thank the participating companies for their collaboration. The 
research was funded by EASE Industrial Excellence Center for Embedded 
Applications Software Engineering (http://ease.cs.lth.se).

\section*{References}
\bibliographystyle{elsarticle-harv}
\bibliography{references}

\end{document}